\RequirePackage{ifpdf}
\ifpdf % We are running pdfTeX in pdf mode
\documentclass[pdftex]{sigma}
\else
\documentclass{sigma}
\fi

\def\cA{{\cal A}}
\def\cD{{\cal D}}

\def\cH{{\cal H}}             % Hilbert space
\def\cN{{\cal N}}

\def\cH{{\cal H}}

\newcommand{\R}{\mathbb{R}}
\def\d{\partial}
\def\bra{\langle}
\def\ket{\rangle}
\def\lr{{\rm L}^2({\R^d})}
\def\dk#1#2{\frac{ d^{#2}{#1} }{ (2\pi)^{#2} }} % invariant measure in FT

\begin{document}
\allowdisplaybreaks

\renewcommand{\PaperNumber}{046}

\FirstPageHeading

\ShortArticleName{Scale-Dependent Functions, Stochastic
Quantization and Renormalization}

\ArticleName{Scale-Dependent Functions, Stochastic Quantization\\
and Renormalization}

\Author{Mikhail V. ALTAISKY~$^{\dag\ddag}$}

\AuthorNameForHeading{M.V. Altaisky}

\Address{$^\dag$ Joint Institute for Nuclear Research, Dubna, 141980 Russia}
\URLaddressD{\url{http://lrb.jinr.ru/people/altaisky/MVAltaiskyE.html}}

\Address{$^\ddag$~Space Research Institute RAS, Profsoyuznaya 84/32, Moscow, 117997
Russia}
\EmailD{\href{mailto:altaisky@mx.iki.rssi.ru}{altaisky@mx.iki.rssi.ru}}

\ArticleDates{Received November 25, 2005, in f\/inal form April
07, 2006; Published online April 24, 2006}

\Abstract{We consider a possibility to unify the methods of
regularization, such as the renormalization group method,
stochastic quantization etc., by the extension of the standard
f\/ield theory of the square-integrable functions $\phi(b)\in \lr$
to the theory of functions that depend on coordinate $b$ and
resolution $a$. In the simplest case such f\/ield theory turns out
to be a theory of f\/ields $\phi_a(b,\cdot)$ def\/ined on the
af\/f\/ine group $G:x'=ax+b$, $a>0,x,b\in\R^d$, which consists of
dilations and translation of Euclidean space. The f\/ields
$\phi_a(b,\cdot)$ are constructed using the continuous wavelet
transform. The parameters of the theory  can explicitly depend on
the resolution $a$. The proper choice of the scale dependence
$g=g(a)$ makes such theory free of divergences by construction.}


\Classification{37E20;  42C40; 81T16; 81T17}

\section{Introduction}

In many problems of f\/ield theoretic description of
inf\/inite-dimensional systems the continuous description of
f\/ields and propagators faces the problem of inf\/inities in loop
integrals. The underlying physics is that f\/ield theoretic
description includes integration over all microscopic scales
smaller than the size of the system. In the simplest case
small-scale f\/luctuations can be just averaged to get the
classical equations for macroscopic f\/ields. The best known
example is the laminar f\/low hydrodynamics. In more complicated
cases interaction strength (and even type) may depend on the
scale. The examples are: turbulent f\/luid f\/low, critical
phenomena, elementary particles interaction, etc.

For this reason we need a measure of integration that works better
than Euclidean integration does. For some cases such measures are
observed experimentally: these are fractal measure of energy
dissipation in turbulent f\/ield \cite{Benzi1984,ABM1991}, fractal
mass distribution of percolation clusters, etc. For high energies,
when the fractal distribution of the f\/ield or order parameter
can not be measured experimentally due to the uncertainty
principle, the search for a better measure of integration remains
a mathematical problem. Fortunately, all these phenomena share a
common important feature -- the self-similarity, or {\em scaling}.
The percolation processes in nanoelectronics, the turbulent
velocity f\/ield, and the nucleon scattering -- all display the
same picture if being observed at dif\/ferent scales. The
self-similarity has initiated at least two important
regularization methods for f\/ield theory models:
\begin{itemize}\vspace{-2mm}
\itemsep=0pt \item The renormalization group (RG) method that
makes use of substitution of initial f\/ields $\phi(x)\in\lr$ by
the scale-truncated f\/ields
\begin{gather}
\phi_{\left(\frac{2\pi}{\Lambda}\right)}(x) = \int_{|k|\le\Lambda}
e^{-\imath k x} \tilde \phi(k) \dk{k}{d}, \label{trunc}
\end{gather}
makes the coupling constants dependent on the cut-of\/f momentum
$\Lambda$, and requires that the f\/inal physical results should
be independent of the introduced scale
\[
\Lambda\d_{\Lambda}({\rm Physical\ quantities})=0.
\]
\item Use of random processes that are self-similar by
construction  for the regularization of quantum f\/ield theory
(QFT) models. This is known as {\em stochastic quantization}.
\vspace{-1mm}
\end{itemize}

In present paper we summarize the both methods by introducing the
f\/ields $\phi_a(b)$ that expli\-citly depend on both position $b$
and the scale (resolution) $a$. If $\phi(x)$ is considered as the
wave function of a quantum particle, then the normalization $\int
|\phi_a(b)|^2 d\mu(a,b)=1$, where $d\mu(a,b)$ is the appropriate
Haar measure on af\/f\/ine group, expresses the fact that the
probability of locating the particle anywhere in space $-\infty <
b < \infty$ changing the resolution $0<a<\infty$ is exactly one.
Thus, the RG symmetry related to the change of the scale $a$
extends the symmetry of the theory by allowing the change of
parameters with scale. Technically, construction of the 
scale-dependent f\/ields $\phi_a(b)$ is performed by using continuous
wavelet transform (CWT).

It should be noted that since f\/ield theoretic calculations are
usually performed with the help of the Fourier transform the
cut-of\/f momentum $\Lambda$ corresponds to the minimal coordinate
scale $L=\frac{2\pi}{\Lambda}$. However, the Fourier transform is
not localized in coordinate space. For this reason we denote the
f\/ields obtained from $\phi(x)$ by truncation of the Fourier
harmonics with $|k|>\frac{2\pi}{L}$ \eqref{trunc} as
$\phi_{(L)}(x)$; in contrast to it the localized view of a
function $\phi$ at the position $x$ and the scale $a$ is denoted
as $\phi_a(x)$. The mathematical meaning of the latter will be
explained hereafter in terms of wavelet transform. The same thing
concerns the parameters of f\/ield theory models -- masses,
coupling constants, etc. In the standard approach they are
dependent on the cut-of\/f momentum and it is tacitly understood
that the experimental dependence of these parameters on the
squared transferred momentum $Q^2$ is equivalent to the dependence
on the cut-of\/f. In our wavelet-based approach the $g=g(a)$ may
be the coupling constant of the given f\/ield $\phi_a(b)$ of the
given scale~$a$, rather than the ef\/fective coupling constant for
the harmonics of
 all scales up to $a$. For this reason dif\/ferent notations are used hereafter in the Fourier and the wavelet
representations of f\/ield theories.

\section{Wavelets and scale-dependent functions}

\subsection{What is wavelet transform}
Wavelet transform entered mathematical physics from geophysical
applications \cite{Zimin1981,GGM1984} as a prefe\-rab\-le
alternative to the Fourier transform in the case when localization
in both position and momentum is simultaneously required. Formally
the Fourier transform and the wavelet transform are on the same
footing: both are decomposition of a function with respect to the
representations of a Lie group. However, the former uses Abelian
group of translations, the latter uses non-Abelian group of
af\/f\/ine transformations
\[
x'=aR(\theta)x+b,\qquad R(\theta) \in SO_d, \quad x,x',b \in \R^d.
\]
Let us remind the general formalism. For a locally compact Lie
group $G$ acting transitively on the Hilbert space $\cH$ it is
possible to decompose vectors $\phi\in\cH$ with respect to the
square-integrable representations $U(g)$ of the group $G$
\cite{Carey1976,DM1976}:
\begin{gather}
|\phi\ket = C_\psi^{-1} \int_{g\in G} |U(g)\psi\ket
d\mu(g)\bra\psi U^*(g)|\phi\ket, \label{pu}
\end{gather}
where $d\mu(g)$ is the left-invariant Haar measure on $G$. The
normalization constant $C_\psi$ is determined by the norm of the
action of $U(g)$ on the f\/iducial vector $\psi\in\cH$, i.e.\ any
$\psi\!\in\!\cH$ that satisf\/ies the admissibility condition
\[
C_\psi = \|\psi \|^{-2} \int_{g\in G} |\bra\psi|U(g)\psi \ket|^2
d\mu(g) <\infty,
\]
can be used as a basis of wavelet decomposition \eqref{pu}.

Hereafter we assume the basic wavelet $\psi$ is invariant under
$SO_d$ rotations $\psi(\boldsymbol{x})=\psi(|\boldsymbol{x}|)$ and
drop the angular part of the measure for simplicity
($R(\theta)\equiv1$). After this simplifying assumption, the
left-invariant Haar measure on af\/f\/ine group is
$d\mu(a,\boldsymbol{b})=\frac{da d^d\boldsymbol{b}}{a^{d+1}}$. The
representation $U(g)$ induced by a basic wavelet $\psi(x)$  is
\begin{gather}
g:\boldsymbol{x}'=a\boldsymbol{x}+\boldsymbol{b},\qquad
U(g)\psi(\boldsymbol{x})=a^{-d/2}\psi\left( \frac{\boldsymbol{x}-
\boldsymbol{b}}{a}\right). \label{ag}
\end{gather}
Therefore, in the Hilbert space of square-integrable functions
$\lr$, with the scalar product $(f,g):=\int \bar f(x) g(x) dx$, a
function $\phi\in\lr$ can be decomposed with respect to the
representations of af\/f\/ine group \eqref{ag}:
\begin{gather}
\phi(x)= C_\psi^{-1}\int
a^{-d/2}\psi\left(\frac{x-b}{a}\right)\phi_a(b)
\frac{dad^db}{a^{d+1}}. \label{iwt}
\end{gather}
The coef\/f\/icients of this decomposition are
\begin{gather}
\phi_a(b)= \int
a^{-d/2}\bar\psi\left(\frac{x-b}{a}\right)\phi(x)d^dx. \label{dwt}
\end{gather}
The $d$-dimensional bold vector notations are dropped hereafter
where it does not lead to a~confusion, and the basic wavelet
$\psi(x)$ is assumed to be isotropic. Here we use normalization
for the rotationally invariant wavelets
\[
C_\psi = \int \frac{|\tilde \psi(k)|^2}{S_d|k|^d}d^dk,
\]
where the area of the unit sphere in $d$ dimensions $S_d$ has come
from rotation symmetry.

\subsection{Scale-dependent functions}
Up to this point we considered only a projection of a $\phi\in\lr$
functions onto the basis constructed of shifts and dilations of
the basic wavelet $\psi(x)$. If we substitute the inverse wavelet
transform \eqref{iwt} into any physical theory we have started
with we will reproduce it identically. However, the wavelet
coef\/f\/icients $\phi_a(b)$ may have their own physical meaning.
The convolution~\eqref{dwt} can be considered as a ``microscope''
that scrutinizes a f\/ield or a signal $\phi(x)$ at a point $b$ at
dif\/ferent resolutions. In this sense the function $\psi$ can be
considered as an apparatus function of the measuring device.

The most known case of such theory is the measurement of the
turbulent velocity f\/ield: the mean velocity and the PDF of
f\/luctuations of dif\/ferent scales should not be the same
\cite{Benzi1984,Frish1995}. Experimentally averaging of the
f\/ield in a physical volume $L^d$ is usually described as
\[
v_{L}= L^{-d}\int_{L^d} K(x-y) v(y) d^dy,
\]
where $K$ is some averaging kernel. But the ``no-scale'' f\/ield
$v(y)$ may not exist at all: consider a mean velocity of
f\/luctuations at a scale $L$ less than a mean free path of the
particle.

That is why the wavelet coef\/f\/icients $\phi_a(b)$ may have
their own operational meaning, even if the ``no-scale'' function
$\phi(x)$ does not exist. In this case a sum of all f\/luctuations
with scales equal or greater than a given scale $L$ can be
def\/ined \cite{Alt1999}:
\[
\phi_{(L)}(x) = \frac{2}{C_\psi} \int_L^\infty
a^{-d/2}\psi\left(\frac{x-b}{a}\right)\phi_a(b)
\frac{dad^db}{a^{d+1}}.
\]
This is a close analogy to the Wilson's RG approach to critical
phenomena \cite{Wilson1971b,WK1974,Wilson1983}, where
a~magnetization (or velocity, or other order parameter) of scale
$L$ is taken in the form \eqref{trunc}.

\subsection{Wavelets and random processes}
If a function to be analyzed by continuous wavelet is a random
function, its wavelet transform is also a random function. So,
instead of the usual space of the random functions
$f(x,\cdot)\in(\Omega, \cA, P)$, where $f(x,\omega)\in \lr$ for
each given realization $\omega$ of the random process, we can go
to the multiscale representation provided by the continuous
wavelet transform \eqref{dwt}:
\[
W_\psi(a,\boldsymbol{b},\cdot) = \int |a|^{-\frac{d}{2}}
\overline{\psi\left(\frac{\boldsymbol{x}-\boldsymbol{b}}{a}\right)}f(\boldsymbol{x},\cdot)d^dx.
\]

The inverse wavelet transform
\[
f(\boldsymbol{x},\cdot) = C_\psi^{-1} \int |a|^{-\frac{d}{2}}
\psi\left( \frac{\boldsymbol{x}-\boldsymbol{b}}{a}\right)
W_\psi(a,\boldsymbol{b},\cdot) \frac{dad\boldsymbol{b}}{a^{d+1}}
\]
reconstructs the common random process as a sum of its scale
components, i.e.\ projections onto dif\/ferent resolution spaces.

The use of the scale components instead of the original stochastic
process provides extra analytical f\/lexibility of the method:
there exist more than one set of random functions
$W(a,\boldsymbol{b},\cdot)$ the images of which  have coinciding
correlation functions in the space of $f(x,\cdot)$. It is easy to
check that the random process generated by wavelet
coef\/f\/icients having in $(a,k)$ space the correlation function
\[
\bra \widetilde W(a_1,k_1) \widetilde W(a_2,k_2)\ket = C_\psi
(2\pi)^{d} \delta^d(k_1+k_2) a_1^{d+1} \delta(a_1-a_2) D_0
\]
has the same correlation function as white noise has:
\begin{gather*}
\bra f(x_1) f(x_2) \ket = D_0 \delta^d(x_1-x_2), \\
\bra {\tilde f}(k_1) \tilde f(k_2)\ket = (2\pi)^d D_0 \delta^d(k_1+k_2), \\
\nonumber \bra \widetilde W_\psi(a_1,k_1) \widetilde
W_\psi(a_2,k_2)\ket =
 (2\pi)^d D_0 \delta^d(k_1+k_2)
(a_1 a_2)^{d/2} \overline{\tilde \psi(a_1 k_1) \tilde\psi(a_2
k_2)}.
\end{gather*}
Therefore, the space of scale-dependent random functions is just
richer than the ordinary space of random functions: since we have
an extra scale argument $a$ here, we can play with the PDF and
correlations of random functions $\phi_a(b,\cdot)$ to achieve
required properties in ordinary space, holding at the same time
some other limitations, say, on singular behavior.

\section{Stochastic quantization}
\subsection[Parisi-Wu stochastic quantization]{Parisi--Wu stochastic quantization}
Quantum f\/ield theory, as is understood in technical sense, is a
tool to calculate  vacuum expectation values (v.e.v.) of the
f\/ield operator products $\bra\phi(x_1)\ldots\phi(x_n)\ket$ --
the Green functions. Usually it is done in the functional integral
formalism: if the action of the f\/ield $\phi$  $S[\phi]$ is
known, the v.e.v.\ can be derived by taking functional derivatives
of the generating functional
\begin{gather}
W[J(x)] = \int \cD\phi e^{\imath S[\phi] + \imath \int J(x)
\phi(x) dx} \label{fi1}
\end{gather}
with respect to the formal source $J(x)$. Unfortunately,
inf\/inite perturbation series obtained by such dif\/ferentiation
often give inf\/inite results and require regularization.

However, there are alternatives to this method. Since any quantum
system interacts with environment, it can not be in a pure state,
and averaging over all f\/ield conf\/igurations should be
performed taking into account averaging over the states of
environment. This can be done by considering the system
environment as a thermostat, described by a Gaussian random noise;
the desired v.e.v.\ state will be a result of relaxation to the
thermodynamic limit. Thus, instead of the Feynman path integral
\eqref{fi1}, one can operate with stochastic dif\/ferential
equations. Because of equivalence between Euclidean QFT and the
stochastic problems, the stochastic methods have found wide
applications in both numeric and analytic solution of QFT
problems, especially in regularization. Among those, the method of
stochastic quantization of gauge f\/ields, proposed by G.~Parisi
and Y.~Wu \cite{PW1981}, have been attracting attention for more
than 20 years. The idea of the method is as follows. Let
$S_E[\phi]$ be the action Euclidean f\/ield theory in $\R^d$.
Instead of direct calculation of the Green functions from the
generation functional of the f\/ield theory, it is possible to
introduce a {\em fictitious time} $\tau$, make the quantum
f\/ields into stochastic f\/ields $\phi(x) \to \phi(x,\tau)$,
$x\in\R^d$, $\tau\in\R $ and evaluate the moments $\langle
\phi(x_1,\tau_1) \ldots \phi(x_m,\tau_m) \rangle$ by averaging
over a random process $\phi(x,\tau,\cdot)$ governed by the
Langevin equation
\begin{gather}
\frac{\d\phi(x,\tau)}{\d\tau} + \frac{\delta
S}{\delta\phi(x,\tau)} = \eta(x,\tau). \label{le}
\end{gather}
The Gaussian random force is $\delta$-correlated in both the
$\R^d$ coordinate and the f\/ictitious time $\tau$:
\begin{gather}
\langle \eta(x,\tau)\eta(x',\tau') \rangle = 2 D_0
\delta(x-x')\delta(\tau-\tau'),\qquad \langle
\eta(x,\tau)\rangle=0. \label{wn1}
\end{gather}
The physical Green functions are obtained by taking the steady
state limit:
\[
G(x_1,\ldots,x_m) = \lim_{\tau\to\infty} \langle \phi(x_1,\tau)
\ldots \phi(x_m,\tau) \rangle.
\]

The stochastic quantization method is much preferable to ordinary
methods for it does not have problems with gauge f\/ixing and does
not require incorporation of higher derivatives as continuous
regularization methods do.

Concerning the renormalization of stochastically quantized theory,
the matter practically comes out to the renormalization of the
Langevin equation \eqref{le} in $(d+1)$ dimensions instead of
renormalization of the original theory with the action functional
$S[\phi]$ in $d$ dimensions. The renormalization of the Langevin
equation is usually done by constructing the characteristic
functional
\[
Z[J]= \left\langle{ \exp\left(\int d^d x d\tau J(x,\tau)
\phi(x,\tau) \right) }\right\rangle_{\rm solutions},
\]
where the statistical averaging $\bra\cdots\ket$ is taken over all
solutions of the Langevin equation \eqref{le}. The summation over
all solutions in functional integral formalism is achieved by
expressing the functional $\delta$-function as a functional
integral over an imaginary auxiliary f\/ield $\hat\phi$:
\begin{gather}
\delta \left(\frac{\d\phi(x,\tau)}{\d\tau} + \frac{\delta
S}{\delta\phi(x,\tau)} -\eta(x,\tau) \right) \sim \int \cD
\hat\phi \exp \hat\phi \left( \frac{\d\phi(x,\tau)}{\d\tau} +
\frac{\delta S}{\delta\phi(x,\tau)} -\eta(x,\tau) \right).
\label{gfle}
\end{gather}
This is so-called Martin--Siggia--Rose f\/ield-doubling formalism
\cite{MSR1973}. The resulting theory has the form of a f\/ield
theory with constraints in $(d+1)$ dimensions. Under the change of
the cut-of\/f scale all f\/ields, coordinates and parameters in
such a theory are transformed according to their canonical
dimensions.

The renormalization of such a theory has much in common with the
renormalization of gauge theories: for the constraint \eqref{gfle}
is similar to gauge f\/ixing, and the functional determinant
arising from the this term under the change of variables results
in ghost f\/ields and BRST symmetry. The details of
renormalization of stochastically quantized theories can be found
e.g.\ in \cite{ZJ1986,Namiki1992}.

\subsection{Wavelet-based stochastic quantization}

Taking into account the above mentioned f\/lexibility of
stochastic processes def\/ined in the space of wavelet
coef\/f\/icients, it appears to be attractive to use
wavelet-def\/ined noise for stochastic quantization. Applying the
wavelet transform (in spatial coordinates, but not in f\/ictitious
time)
\[
\phi_a(b,\cdot) = \int a^{-d}
\overline{\psi\left(\frac{x-b}{a}\right)}\phi(x,\cdot)d^dx
\]
to the f\/ields and the random force in the Langevin equation
\eqref{le}, we get the possibility to change the white noise
\eqref{wn1} into a scale-dependent random force
\begin{gather}
\langle \tilde \eta_{a_1}(k_1,\tau_1) \tilde
\eta_{a_2}(k_2,\tau_2) \rangle = C_\psi (2\pi)^d \delta^d(k_1+k_2)
\delta(\tau_1-\tau_2) a_1 \delta(a_1-a_2) D(a_1,k_1). \label{gns1}
\end{gather}
(Here and after we use CWT in $L^1$ norm instead of $L^2$.)

In the case when the spectral density of the random force  is a
constant $D(a_1,k_1)=D_0$, the inverse wavelet transform
\begin{gather}
\phi(x)  = \frac{2}{C_\psi} \int_0^\infty \frac{da}{a} \int
\dk{k}{d} \frac{d\omega}{2\pi} \exp(\imath(kx-\omega\tau)) \tilde
\psi(ak) \phi_a(k,\omega) \label{zt}
\end{gather}
drives the process \eqref{gns1} into the white noise \eqref{wn1}.

In the case of arbitrary functions $\phi_a(x,\cdot)$ we have more
possibilities. In particular, we can def\/ine a narrow band
forcing that acts at a single scale
\begin{gather}
D(a,k) = a_0 \delta(a-a_0) D_0. \label{sb}
\end{gather}

The contribution of the scales with the wave vectors apart from
the the typical scale $a_0^{-1}$ is  suppressed by rapidly
vanishing wings of the compactly supported wavelet
$\tilde\psi(k)$.

Here we present two examples of the divergence free stochastic
perturbation expansion: ({\em i})~the scalar f\/ield theory
$\phi^3$, ({\em ii}) the non-Abelian gauge f\/ield  theory.

\subsection[Scalar field theory]{Scalar f\/ield theory}
\begin{figure}[h]
\centering\includegraphics[width=6cm]{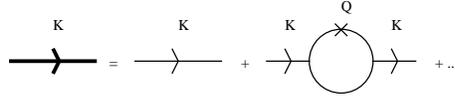}
\caption{Diagram expansion of the stochastic Green function in
$\phi^3$-model.} \label{gf:pic}
\end{figure}

Let us turn to the stochastic quantization of the $\phi^3$-theory
with the  scale-dependent noise \cite{Alt2002G24}. The Euclidean
action of the $\phi^3$-theory is:
\[
S_E[\phi(x)]= \int d^d x \left[\frac{1}{2} (\d\phi)^2 +
\frac{m^2}{2}\phi^2 + \frac{\lambda}{3!}\phi^3 \right].
\]
The corresponding Langevin equation is written as
\begin{gather}
\frac{\d\phi(x,\tau)}{\d\tau} + \left( -\Delta \phi + m^2\phi +
\frac{\lambda}{2!} \phi^2 \right) = \eta(x,\tau). \label{le3}
\end{gather}
Substituting the scale components in representation \eqref{zt}
into the Langevin equation \eqref{le3} we get the integral
equation for the stochastic f\/ields:
\begin{gather}
(-\imath\omega + k^2 + m^2)\phi_a(k,\omega) =  \eta_a(k,\omega)
-\frac{\lambda}{2}  \overline{\tilde\psi(ak)}
\left(\frac{2}{C_\psi}\right)^2
\int\dk{k_1}{d}\frac{d\omega_1}{2\pi}\frac{da_1}{a_1}\frac{da_2}{a_2},\nonumber\\
\tilde\psi(a_1k_1) \tilde\psi(a_2(k-k_1)) \phi_{a_1}(k_1,\omega_1)
\phi_{a_2}(k-k_1,\omega-\omega_1) . \label{phi3w}
\end{gather}
Starting from the zero-th order approximation $\phi_0 = G_0 \eta$,
with the bare Green function $G_0(k,\omega) = 1/(-\imath\omega +
k^2 + m^2)$, and iterating the integral equation \eqref{phi3w}, we
get the one-loop correction to the stochastic Green function:
\begin{gather}
G(k,\omega) =  G_0(k,\omega) \nonumber\\
\phantom{G(k,\omega) =}{}+ \lambda^2 G_0^2(k,\omega) \int
\dk{q}{d}\frac{d\Omega}{2\pi}  2\Delta(q) |G_0(q,\Omega)|^2
G_0(k-q,\omega-\Omega) +\cdots, \label{phi3G1}
\end{gather}
where $\Delta(k)$ is the scale-averaged ef\/fective force
correlator
\[
\Delta(k) \equiv  \frac{2}{C_\psi} \int_0^\infty \frac{da}{a}
|\tilde\psi(ak)|^2 D(a,k).
\]
In the same way  other stochastic moments are evaluated. Thus, the
common stochastic diagram technique is reproduced, but with the
scale-dependent random force \eqref{gns1} instead of the standard
one~\eqref{wn1}. The diagrams corresponding to the stochastic
Green function decomposition~\eqref{phi3G1} are shown in
Fig.~\ref{gf:pic}.

It can be seen that for a single-band forcing \eqref{sb} and a
suitably chosen wavelet the loop divergences are suppressed. For
instance, use of the Mexican hat wavelet
\[
\tilde \psi(k) = (2\pi)^{d/2} (-\imath k)^2 \exp(-k^2/2), \qquad
C_\psi = (2\pi)^d,
\]
and the single-band random force \eqref{sb} gives the ef\/fective
force correlator
\begin{gather}
\Delta(q) = (a_0q)^4 e^{-(a_0q)^2}D_0 \label{efc2}.
\end{gather}
The loop integrals, taken with this ef\/fective force
\eqref{efc2}, can be easily seen to be free of ultraviolet
divergences at $q\to\infty$:
\[
G_2(k,\omega)  =  G_0^2(k,\omega) \int \dk{q}{d} 2\Delta(q)
\int_{-\infty}^\infty \frac{d\Omega}{2\pi} \frac{1}{\Omega^2 +
(q^2+m^2)^2} \frac{1}{-\imath(\omega-\Omega) + (k-q)^2+m^2}.
\]

\subsection{Non-Abelian gauge theory}
The Euclidean action of a non-Abelian gauge f\/ield is given by
\begin{gather}
S[A] = \frac{1}{4} \int d^d x F_{\mu\nu}^a(x)F_{\mu\nu}^a(x),
\nonumber \\
 F_{\mu\nu}^a(x) = \d_\mu A_\nu^a(x) -\d_\nu A_\mu^a(x)
+g f^{abc}A_\mu^b(x) A_\nu^c(x),\label{ym1}
\end{gather}
where $f^{abs}$ are the structure constants of the gauge group,
$g$ is the coupling constant. The Langevin equation for the
stochastic quantization of gauge theory \eqref{ym1} can be written
as
\begin{gather}
\frac{\d A_\mu^a(x,\tau)}{\d\tau} + \bigl( -\delta_{\mu\nu}\d^2 +
\d_\mu \d_\nu \bigr)A_\nu^a(x,\tau)  = \eta_\mu^a(x,\tau) +
U_\mu^a(x,\tau), \label{le2}
\end{gather}
where $\eta_\mu^a(x,\tau)$ is the random force and
$U_\mu^a(x,\tau)$ is the nonlinear interaction term
\[
U[A] = \frac{g}{2}V^0(A,A) + \frac{g^2}{6}W^0(A,A,A).
\]
The two terms standing in the free f\/ield Green function
correspond to the transversal and the longtitudal mode
propagation:
\[
G_{\mu\nu}^{ab}(k) = \frac{T_{\mu\nu}(k)\delta_{ab}}{-\imath\omega
+ k^2}
                   + \frac{L_{\mu\nu}(k)\delta_{ab}}{-\imath\omega}, \qquad
T_{\mu\nu}(k) = \delta_{\mu\nu} - \frac{k_\mu k_\nu}{k^2}, \qquad
L_{\mu\nu}(k)=\frac{k_\mu k_\nu}{k^2}.
\]

Similarly to the scalar f\/ield theory, we can use the
scale-dependent forcing \eqref{gns3} in the Langevin equation
\eqref{le2}. Since there is no dynamic evolution for the
longtitudal modes in the Langevin equation \eqref{le2}, it is
natural to use the transversal scale-dependent random force
\begin{gather}
 \langle \eta_{a_1\mu}^a (k_1,\tau_1) \eta_{\nu a_2}^b(k_2,\tau_2)
\rangle = (2\pi)^d \delta^d(k_1\!+k_2) \delta(\tau_1\!-\tau_2)
T_{\mu\nu}(k_1)
 C_\psi a_1 \delta(a_1\!-a_2) D(a_1,k_1).\!
\label{gns3}
\end{gather}
Let us consider a gluon loop with two cubic vertices. Summing up
over the gauge group indices $ \bigl(\frac{\imath}{2}g\bigr)^2
f^{abc}\delta_{bd}f^{der}\delta_{cr}=
\frac{g^2}{4}\delta_{ae}C_2,$ with $C_2=N$ for $SU_N$ gauge
groups, we can write the gluon loop as a sum of two diagrams --
those with the transversal and the longtitudal stochastic Green
functions:
\begin{gather*}
G_{2\mu\nu}^{ab}(k,\omega)= g^2\delta_{ab}C_2 |G_0(k,\omega)|^2
\sum_{I=T,L}  \int \frac{d\Omega}{2\pi} \dk{q}{d}
N^I(k,\omega,q,\Omega)l^I_{\mu\nu}(k,q) 2 \Delta(q),
\end{gather*}
where
\begin{gather*}
N(k,q) = \left|\frac{1}{-\imath\Omega+q^2}\right|^2
\begin{pmatrix}
\frac{1}{-\imath(\omega-\Omega)+(k-q)^2} \cr
\frac{1}{-\imath(\omega-\Omega)}
\end{pmatrix},
\\
l_{\mu\nu}(k,q) = V_{\mu\kappa\lambda}(k,k-q,q)
T_{\lambda\gamma}(q) V_{\sigma\nu\gamma}(k-q,k,-q)
\begin{pmatrix}
T_{\kappa\sigma}(k-q) \cr L_{\kappa\sigma}(k-q)
\end{pmatrix}.
\end{gather*}
As it can be observed after explicit evaluation of the tensor
structures $l^T_{\mu\nu}$ and $l^L_{\mu\nu}$ using the force
correlator \eqref{gns3}, and integration over $d\Omega$, the
wavelet factor in the ef\/fective force correlator $\Delta(q)$
suppresses the divergences in the case of a narrow-band forcing
\eqref{sb}. The power factor $k^n$ of the basic wavelet $\psi$
that provides $\tilde\psi(0)=0$ and makes the IR behavior softer.
In this respect the wavelet regularization is dif\/ferent from the
continuous regularization  $\int d^dy R_\Lambda(\d^2)
\eta(y,\tau)$, which makes UV behavior softer by the factor
$e^{-\frac{k^2}{\Lambda^2}}$, but do not af\/fect the IR behavior,
see e.g.~\cite{Halpern1993}.

%%%%%%%%%%%%%%%%%%%%%%%%%%%%%%%%%%%%%%%%%%%%%%%%%
\section[Wavelet-based Euclidean field theory]{Wavelet-based Euclidean f\/ield theory}
\subsection{QFT on a Lie group}
Let us consider a Euclidean f\/ield theory determined by a
characteristic functional
\[
W_E[J] = \cN \int \cD\phi \exp\left[-S_E[\phi(x)] + \int  d^dx
J(x)\phi(x) \right]
\]
where $S_E[\phi(x)]$ is Euclidean action, $\cN$ is formal
normalization constant. The (connected) Green  functions
($m$-point cumulative moments) are evaluated as functional
derivatives of the logarithm of generating functional $W_E[J] =
e^{-Z_E[J]}$:
\begin{gather*}
G_m(x_1,\ldots, x_m) \equiv \bra \phi(x_1)\ldots\phi(x_m)\ket = -
\left. \frac{\delta^m}{\delta J(x_1)\cdots\delta
J(x_m)}\right|_{J=0} \ln W_E[J].
%\label{gfm}
\end{gather*}
Applying a formal partition of a unity \eqref{pu} with respect to
a Lie group $G$ we yield a theory with the generating functional
\begin{gather*}
W_g[J(g)] = \cN \int \cD\phi(g) \exp\left[-S_g[\phi(g)] + \int_G
d\mu(g) J(g)\phi(g) \right]
,\qquad g \in G%\label{gfg}
\end{gather*}
and appropriately def\/ined Green functions.

\subsection{Wavelet-based action}
Let us consider the theory of a massive scalar f\/ield with
polynomial interaction
\begin{gather}
S_E[\phi] =\int_{\R^d} d^d x \left[{ \frac{1}{2}(\d_\mu\phi)^2+
\frac{m^2}{2}\phi^2 + \lambda V(\phi)}\right], \label{fe}
\end{gather}
that can be alternatively interpreted as a theory of classical
f\/luctuating f\/ield with the Wiener probability measure $\cD P =
e^{-S_E[\phi]}\cD\phi$. In this case $m^2 = |T-T_c|$ is the
deviation from critical temperature and $\lambda$ is the
f\/luctuation interaction strength.

To introduce the scale-dependent f\/ields $\phi_a(b)$ into the
action we can just substitute the f\/ields~$\phi$ into \eqref{fe}
using wavelet transform \eqref{iwt}. For convenience and recording
purposes, we rewrite continuous wavelet transform (\ref{dwt}),
(\ref{iwt}) using the $L^1$ norm
\begin{gather}
\phi_a(b) = \int \frac{1}{a^d}
\overline{\psi\left(\frac{x-b}{a}\right)}\phi(x)d^dx, \nonumber\\
%\label{dwt1}\\
\phi(x) = \frac{2}{C_\psi} \int_0^\infty \frac{da}{a}\int_{\R^d}
d^db \phi_a(b)  \frac{1}{a^d} \psi\left(
\frac{\boldsymbol{x}-\boldsymbol{b}}{a}\right),
\label{iwt1} \\
 C_\psi = S_d^{-1} \int \frac{|\tilde\psi(\boldsymbol{k})|^2}{|k|^d}d^dk
= 2\int_0^\infty \frac{|\tilde\psi(a\boldsymbol{k})|^2}{a}da <
\infty.\nonumber
\end{gather}
This provides the f\/ields $\phi(x)=\bra x|\phi \ket$ and
$\phi_a(x)=\bra a,x;\psi|\phi \ket$ with the same physical
dimension. The generating functional for the scale-dependent
functions $\phi_a(b)$ is
\[
W[J_a(b)] = \int \cD \phi_a(b) \left(-S_{EW}[\phi_a(b)] + \int
J_a(b) \phi_a(b) \frac{dad^db}{a}\right),
\]
where $J_a(b)$ is a formal source (``external force''), which
corresponds to the f\/luctuations of given scale $a$ localized
near a given point $b$. The corresponding Green functions a given
by
\[
G(a_1,x_1;\ldots;a_n,x_n) = -\left. \frac{\delta \ln
W[J_a(x)]}{\delta J_{a_1}(x_1)\delta J_{a_n}(x_n) }\right|_{J=0} =
\bra \phi_{a_1}(x_1)\ldots \phi_{a_n}(x_n)\ket.
\]

Let us consider a theory with polynomial interaction
$V(\phi)=\frac{g}{n!}\phi^n$
\begin{gather}
S_{EW} =  \int -\frac{1}{2} \phi_{a_1}(b_1)D(a_1,a_2;b_1-b_2)
\phi_{a_2}(b_2) + \frac{m^2}{2} |\phi_a(b)|^2
\nonumber \\
\phantom{S_{EW} =}{} +\frac{g}{n!} V^{a_1\ldots a_n}_{b_1\ldots
b_n} \phi_{a_1}(b_1)\cdots
\phi_{a_n}(b_n),\label{wna}\\
V^{a_1\ldots a_n}_{b_1\ldots b_n} = \int d^dx V(x_1,\ldots,x_n)
\prod_{i=1}^n \frac{1}{a_i^d} \psi\left(\frac{x-b_i}{a_i}\right),
\label{wnp}
\end{gather}
where integration over all pairs of matching indices $\frac{da_i
d^db_i}{C_\psi a_i}$ is assumed. The inverse propagator is the
wavelet image of ($-\d^2+m^2$), that is
\[
D(a_1,a_2,k) =
\overline{\tilde\psi(a_1k)}(k^2+m^2)\tilde\psi(a_2k).
\]
As it is seen from \eqref{wnp}, the local interaction term
$\phi^n$ becomes nonlocal  after the application of wavelet
transform \eqref{iwt1}. Of course, since the wavelet transform
provides the partition of a unity, the integration over all scale
arguments in the theory \eqref{wna} drives it back to the original
f\/ield theory with polynomial interactions, which is UV
divergent. However, using the Wilson's RG ideas we shall show how
the nonlocal wavelet theory~\eqref{wna} can be made into a local
one, which has UV f\/inite behavior.

Let us suppose that we manage to make the wavelet theory with
nonlocal interaction \eqref{wnp} into a local one with the
coupling constant explicitly dependent on scale $g=g(a)$. The
simplest case of the fourth power interaction of this type is
\begin{gather}
V_{int}[\phi] =\frac{1}{C_\psi} \int \frac{g(a)}{4!} \phi^4_a(b)
\frac{dadb}{a}, \qquad  g(a)\sim a^\nu. \label{vint}
\end{gather}
The one-loop order contribution to the Green function $G_2$ in the
theory with local interaction~\eqref{vint} can be easily
evaluated~\cite{ffp4} by integration over a scalar variable $z =
ak$:
\begin{gather*}
\int \frac{a^\nu  |\tilde\psi(ak)|^2
}{k^2+m^2}\dk{k}{d}\frac{da}{a} = C_\psi^{(\nu)}  \int \dk{k}{d}
\frac{k^{-\nu}}{k^2+m^2},
\end{gather*}
where
\[
C_\psi^{(\nu)}=C_\psi^{-1}\int |\tilde\psi(z)|^2 z^{\nu-1}dz.
\]
Therefore, there are no UV divergences for $\nu>d-2$. However, the
positive values of $\nu$ mean that the interaction strengths at
large scales and diminishes at small. (This is a kind of
asymptotically free theory that is hardly appropriate say to
magnetic systems.)

%%%%%%%%%%%%%%%%%%%%%%%%%%%%
We have to admit that the idea of using wavelets for
regularization and blocking of degrees of freedom has been
considered by G.~Battle \cite{Battle-book} and P.~Federbush
\cite{Federbush1981} in lattice settings. The idea was to sum up
the discrete wavelet expansion of a function starting from a large
infra-red scale (set to unity) up to the inf\/initesimally small
scale instead of integrating in Fourier space. The expansion was
taken in the form
\begin{gather}
\phi(x) = \sum_k \alpha_k u_k(x), \qquad u_k(x) =
\frac{1}{\sqrt{-\Delta + m^2}} \Psi_k(x), \label{federbush}
\end{gather}
where $k=(n,m,s)$ is a multiindex that incorporates translations
and binary dilations on the lattice as well as internal degrees of
freedom of the f\/ield $\phi$, $\Psi_k(x)$ forms an orthogonal
basis in~$\lr$. It have been understood then that dif\/ferent
subgrids involved in the wavelet expansion~\eqref{federbush}
represent dif\/ferent independent degrees of freedom
\cite{Federbush1986}, rather than being just dif\/ferent
approximation of the same f\/ield like that in Kadanof\/f block
expansion. The corresponding diagram technique, called ``wavelet
diagrams'' and originally proposed by Federbush, can be found in
Battle's book \cite{Battle-book}.

For a particular type of wavelet basis, the Lemari\'e wavelets, it
was shown that due to the big number of vanishing momenta of the
basic wavelets $\Psi_k$ the correlations between dif\/ferent block
variables $\bra\alpha_k \alpha_{k'} \ket$ decrease as a high
inverse power of the separation between blocks. As a consequence
of this a f\/ield theory $S[\phi(x)]$ becomes UV-f\/inite without
any renormalization. However the decoupling of interaction between
dif\/ferent scales seems to have quite general nature and that is
the reason we exploit this idea in present paper in terms of {\it
continuous} wavelet transform without using any lattices.

Now we will try to show how the local interaction of type
\eqref{vint} can emerge. Following
K.~Wilson~\cite{WK1974,Wilson1983} we consider a system of spins
with spacing $L_0$. The idea of RG applied to a~spin system is
that the interaction of spins separated by a distance $L_0$, can
be substituted by interaction of blocks (of $2^d$ spins in each)
of size $2L_0$ with the high-frequency details absorbed into the
interaction constants. The same procedure is then applied to the
blocks of $2L_0$ size, blocks of $4L_0$ size etc.\ -- this is
so-called Kadanof\/f blocking procedure. On each step of the
blocking procedure the transformation from the scale $L$ to $2L$
is done by integrating over the high frequency degrees of freedom
$K>\frac{2\pi}{L}$ with appropriate adjustment of the coupling
constants.

In continuous limit, i.e.\ in Landau theory of ferromagnetism,
when spins are averaged into the mean magnetization $\phi(x)$, the
energy of the spin system is given by the Ginzburg--Landau free
energy functional
\begin{gather}
F_E[\phi] = \int d^dx \left[ \frac{1}{2} (\d\phi)^2 + \frac{R}{2}
\phi^2 + \frac{U}{4!}\phi^4 \right], \label{gle}
\end{gather}
where integration is performed over the domain occupied by the
spin system. The f\/ield theory \eqref{gle} gives  divergences in
correlation functions. To cope with it  we remind that $\phi(x)$
is the mean magnetization of a macroscopically big block, i.e.\
one that contains enough spins to make statistical averaging
valid. This means that the f\/ield $\phi$ should be substituted by
one containing only the f\/luctuations of size larger than a
typical scale $L$:
\begin{gather}
\phi(x) \to \phi_{(L)}(x)=\int_{|k|<\frac{2\pi}{L}} e^{-\imath k
x} \tilde\psi(k)\dk{k}{d}. \label{phih}
\end{gather}

So, the original Ginzburg--Landau functional is made into
ef\/fective action for the large-scale f\/ields $\phi_{(L)}$
\begin{gather}
F_L[\phi_{(L)}] = \int d^dx \left[ \frac{1}{2} (\d\phi_{(L)})^2 +
\frac{R_L}{2} \phi^2_{(L)} + \frac{U_L}{4!}\phi^4_{(L)} \right],
\label{glh}
\end{gather}
in which the ef\/fects of small-scale f\/luctuations with $|k|\ge
\frac{2\pi}{L}$ are absorbed into the coupling constants $R_L$ and
$U_L$. It should be noted that the projection \eqref{phih},
performed by f\/iltering out the Fourier harmonics with $|k|\ge
\frac{2\pi}{L}$, is not the only possible type of smoothing: an
exponential or proper-time cutof\/f can be used as well
\cite{Gaite2004}. Similar projection can be obtained if we apply
the wavelet transform with certain given kernel $\phi(x)\to
\phi_a(b)$.

Wilson suggested an elegant way to determine the dependence of
 $R_L$ and $U_L$ on the cutof\/f scale $L$ by averaging over f\/luctuations
in the shell $[L,L+\delta L)$ \cite{WK1974,
Wilson1971b,Wilson1983}. Since we understand the f\/luctuations of
a given scale $\phi_a(b)$ -- now in wavelet notation -- as
physical f\/ields measured at scale $a$, we will reproduce the
Wilson's derivation integrating the scale-dependent free-energy
$F[\phi_a(b)]$ in appropriate limits over the logarithm of scale
$a^{-1}da$. The larger is the scale $L$, the more f\/luctuations
of smaller scales contribute to free energy.

Let $L_0$ be the smallest size of the system -- the distance
between spins in the theory of ferromagnetism. Then, the
Kadanof\/f blocking procedure is a mapping
\begin{gather*}
\{ \psi^{(0)}_k \} \to \{ \psi^{(1)}_k \} \to \{ \psi^{(2)}_k \} \to \cdots,  \\
H^{(0)}(R_0,U_0) \to H^{(1)}(R_1,U_1) \to H^{(2)}(R_2,U_2) \to
\cdots,
\end{gather*}
where $\{ \psi^{(0)}_k \}$ is the  basis for the f\/inest
resolution Hilbert space, with $H^{(0)}(R_0,U_0)$ being the
Hamiltonian acting on this space, $\{ \psi^{(1)}_k \}$ is the
basis of the next coarse-grained space of the resolution $2L_0$,
etc. At each step of the coarse graining process some details are
lost, so any function $\psi^{(1)}_k$ of the basis $\{ \psi^{(1)}_k
\}$ can be expanded in the basis $\{ \psi^{(0)}_k \}$; and
similarly for all next scales:
\[
\psi^{(j)}_i = \sum_k c^{j+1}_{ik} \psi^{(j)}_k,
\]
since $\{ \psi^{(j)}_k \}$ is more detailed than $\{
\psi^{(j+1)}_k \}$.

Expanding the more coarse-grained basic functions in terms of the
less coarse-grained one we obtain the expansion of the parameters
of the coarser Hamiltonian $H^{(j+1)}$ in terms of the f\/iner
Hamiltonian  $H^{(j)}$. Wilson suggested a qualitative way of
doing that. Suppose we know the free energy functional $F_L[\phi]$
of scale $L$ and need to calculate the next coarser functional
$F_{L+\delta L}[\phi]$ of scale $L+\delta L$. Then, we have to add
f\/luctuations of all scales within the range $[L,L+\delta L)$ to
the original theory and integrate over those f\/luctuations. If
$\psi(x)$ is a normalized basic function of scale $L$, e.g.\ a
wave packet with $k\approx \frac{2\pi}{L}$, such that
\begin{gather}
\int d^dx |\psi(x)|^2 = 1,\qquad \int d^dx |\d\psi(x)|^2 \simeq
L^{-2}, \label{l2c}
\end{gather}
we just make $\phi(x) \to \phi(x) + c \psi(x)$ and integrate over
the scalar amplitude $c$:
\begin{gather}
e^{-F_{L+\delta L}[\phi]} = \int_{-\infty}^{\infty} dm
e^{-F_{L}[\phi+ m \psi]}. \label{adfl}
\end{gather}
Substituting the free energy \eqref{glh} into the condition
\eqref{adfl} and taking into account the conditions~\eqref{l2c} we
get
\begin{gather}
e^{-F_{L+\delta L}[\phi]} = e^{-F_{L}[\phi]}
\int_{-\infty}^{\infty} dm e^{-\frac{m^2}{2} \left(
\frac{1}{L^2}+R_L +\frac{1}{2} U_L\phi^2 \right)}. \label{gife}
\end{gather}
The derivation is not very rigorous, and referring the reader to
the original paper \cite{Wilson1983} for the details, we just have
to state that using the formal properties of Gaussian integration
in \eqref{gife}, up to the numerical factors, we get a logarithmic
contribution to the free energy
\begin{gather}
F_{L+\delta L}[\phi] = F_L[\phi]+ \frac{1}{2} \ln \left(
\frac{1}{L^2}+R_L +\frac{1}{2} U_L\phi^2 \right). \label{lcf}
\end{gather}
To derive the dependence of $R_L$ and $U_L$ on scale $L$ two more
tricks were used by Wilson: (i)~the logarithm in \eqref{lcf} is
decomposed into a Taylor series; (ii)~integration over the phase
space volume corresponding to the d.o.f.\ related to the
wavepacket $\psi$ is changed into integration over the coordinate
volume $\delta V$ using the uncertainty principle.

The logarithm is decomposed into the power series only up to the
second order -- to get the forth order in the f\/ields. Up to the
terms that do not depend on $\phi$ we get
\begin{gather}
\frac{1}{2} \ln \left( \frac{1}{L^2}+R_L +\frac{1}{2} U_L\phi^2
\right) =\frac{1}{4}\left( U_L L^2 \phi^2 -\frac{1}{4} L^4 \phi^4
-R_L U_L L^4\phi^2\right) + \cdots. \label{lf1}
\end{gather}

The next trick is to count the degrees of freedom corresponding to
the localized wavepacket~$\psi$. Since the scales are in the range
$[L,L+\delta L)$, the volume in momentum space is $\frac{\delta
p}{2\pi}\sim \frac{\delta L}{L^2}$. The phase space volume, since
$p \approx 2\pi/L$, is
\[
 \delta \Gamma \sim V p^{d-1} \delta p \sim V L^{-1-d} \delta L,
 \]
and, hence, the coordinate volume per one degree of freedom is
\[
\delta V \sim \frac{L^{1+d}}{\delta L}.
\]
Thus the equation \eqref{lf1} can be formally multiplied by
$1=\delta V L^{-1-d}\delta L$ and integrated over the volume
$\delta V$. This results in equations
\begin{gather*}
R_{L+\delta L} = R_L +\frac{1}{2}\left({ L^{1-d}U_L -L^{3-d}R_L U_L} \right)\delta L,\\
U_{L+\delta L} = U_L -\frac{3}{2} L^{3-d}U_L^2 \delta L.
\end{gather*}
See \cite{Z-J1989} for general theory of renormalization in
f\/ield theory and critical phenomena.

%%%%%%%%%%%%%%%%%%%%%%%%%%%%%%%%%%%%%%%%%%%%%%%%%%%%%%%%%%%%%%%%%%%%%%%
The above Wilson's consideration was presented only to show how
integration in the thin shell below the cutof\/f renormalizes the
parameters of the action. More rigorous formalism that accounts
for the change of the action parameters by means of integration
over the inf\/initesimally thin shell $1-\delta l < |q| <1$ in
momentum space, followed by rescaling, consists in constructing
dif\/ferential equations for the variation of the action
functional. It is known as exact renorma\-li\-zation group (ERG). The
ERG equations are rather complicated and can be found elsewhere
\cite{WH1973,WK1974,Kub1998}. If solved numerically, the ERG
equations could provide
 exact scale dependence of the coupling constants on cutof\/f scale in terms of the considered model.
The problems are, however, that (i) it is often impossible to
solve these equations, (ii) the cutof\/f scale is not the same as
the scale of observation, and hence it is not obvious how to
interpret the ERG results; (iii) and the last, but not the least,
that it is not clear how to separate the modes to be integrated
over from those to be renormalized by this integration using the
space of functions that depend only on momentum $\phi(k)$
(\cite{WK1974}, Fig.~11.2). Here we propose an alternative
solution of the mode separation problem: we just extend the space
of functions adding the scale argument explicitly $\phi(\cdot) \to
\phi_a(\cdot)$.

Having the information on the dependence on scale of the
ef\/fective mass $R_L$ and coupling constant $U_L$ we can address
the question, how the ef\/fective theory \eqref{glh} that
comprises nonlocal interactions of all modes larger than $L$ can
be transformed into a theory with the local interactions of scales
of the type \eqref{vint}. To do this we approximate the original
Ginzburg--Landau action $F$, by a new action
\begin{gather}
F_E[\phi] = \int \frac{dad^db}{a} \left[ -\frac{1}{2}
\phi_{a_1}(b_1)D\phi_{a_2}(b_2) + \frac{m^2(a)}{2} |\phi_a(b)|^2 +
\frac{g(a)}{4!}|\phi_a(b)|^4 \right]. \label{sa}
\end{gather}
That  meets physical requirements by construction: for a free
f\/ield ($g(a)=0$) it coincides with the original free action; for
the interacting f\/ields, it just satisf\/ies the Wilson's
assumption that {\em only the f\/luctuations of the close scale
interact}: in the case of the action \eqref{sa}, certainly the
equal scales only.

Since the action \eqref{sa} explicitly involves integration in
scale variable $da/a$, we can attribute the dif\/ference in the
free energies \eqref{lf1} to interaction of the appropriate
``renormalized'' modes, governed by the equation \eqref{sa}, in
the shell $[L,L+\delta L)$.

The $n>2$ the polynomial interactions are nonlocal in wavelet
space. To derive their interaction constant dependence on scale,
we use the physical assumptions that only the f\/luctuations of
close scales can directly interact to each other. So, in the
self-interaction of the f\/ield $\phi$
\[
U_L\int_{L} \phi(x)^n d^dx = U_L\int_{L} V_{b_1\ldots
b_n}^{a_1\ldots a_n} \phi_{a_1}(b_1) \cdots \phi_{a_n}(b_n)
\prod_{i=1}^n\frac{da_id^db_i}{a_i}
\]
the terms with $|\ln a_i-\ln a_j|\ll 1$ should dominate, from
where we can postulate that
\begin{gather}
U_L\int_{L} V_{b_1\ldots b_n}^{a_1\ldots a_n} \phi_{a_1}(b_1)
\cdots \phi_{a_n}(b_n) \prod_{i=1}^n\frac{da_id^db_i}{a_i} =
\int_L \frac{da}{a} \int  g(a) |\phi_a(b)|^n d^db \label{phi-loc}
\end{gather}
for some $g(a)$.

It should be noted that similar idea of orthogonalization of
f\/luctuations of dif\/ferent scales has been already applied to
Ginzburg--Landau model by C.Best using discrete wavelet
transform. However, the Best's paper \cite{Best2000}, having
started from continuous Ginzburg--Landau model, uses {\em
orthogonal wavelets} and suf\/f\/iciently strong assumptions that
f\/luctuations of dif\/ferent scales are $\delta$-correlated in
both position and scale (equation~(4) from~\cite{Best2000}). This
is def\/initely not the case for arbitrary (non-orthogonal)
wavelets and standard assumption of Gaussian nature of
f\/luctuations in ordinary continuous model~\eqref{gle}.
Nevertheless, the Best's idea is numerically convenient and it is
interesting for what type of wavelets such behavior is really
observed.

Let us perform the asymptotic estimation of the behavior of $g(a)$
at the known behavior of~$U_L$. Let us assume that interaction
takes place at the thin shell of scales $[L,L+\delta L)$ and set
$a_1=\cdots =a_n=L$. Since we need the behavior of the local
interaction of the type \eqref{vint}, we set
 $b_1=\cdots =b_n=0$ in the r.h.s.\ of \eqref{phi-loc} to get the upper bound
for $g(a)$. Doing so, we get
\begin{gather*}
V_{0\ldots0}^{a\ldots a} = \int d^dx \frac{1}{a^{nd}}
\psi^n\left(\frac{x}{a}\right) = a^{(1-n)d} \int d^dy \psi^n(y).
\end{gather*}
In Wilson's theory ($n=4$, $d<4$)
\[
 U_L = \frac{1}{\frac{3}{2}\frac{L^{4-d}}{4-d}+{\rm const}}
\stackrel{\scriptsize{L\to\infty}}{\sim} L^{d-4},
\]
and hence
\begin{gather}
g(L) \sim L^{-3d}U_L \sim L^{-2d-4}. \label{glw}
\end{gather}
The important point is that in the theory \eqref{sa} the
observable quantities are the correlators of the scale-dependent
functions $\langle\phi_{a_1}(b_1)\phi_{a_2}(b_2) \rangle$. If we
deal with ``renormalized'' action \eqref{sa}, where only equal
scales interact polynomially, the loop integrals can be easily
evaluated for a f\/ixed value of scale. In the fourth power
interaction theory the one-loop contribution to the Green function
will be
\begin{gather}
\int_{a_{\min}}^{a_{\max}} \frac{da}{a} g(a) \int \dk{k}{d}
\frac{|\tilde\psi(ak)|^2}{k^2+m^2} = \int_{a_{\min}}^{a_{\max}}
\frac{da}{a} g(a) L(a). \label{ar}
\end{gather}
The $k$-dependent integral in the last equation \eqref{ar} can be
explicitly evaluated. As an example, let us present a ``bad'' case
of the Morlet wavelet $\tilde\psi(k)=\exp\left(\imath k z
-\frac{k^2}{2}\right)$, which is not admissible in the sense
$C_\psi=\infty$,
\begin{gather*}
L(a) = S_d \int_0^\infty
\frac{e^{-a^2q^2}q^{d-1}}{q^2+m^2}\dk{q}{d}
\stackrel{d=4,m=1}{\rightarrow}  \frac{-a^4e^{a^2}
Ei(1,a^2)+a^2}{a^2} \frac{a^{2-d}}{2}.
\end{gather*}
For admissible wavelets ($C_\psi<\infty$) the IR behavior is
milder.

Certainly, the integral $\int_{a_{\min}}^{a_{\max}} \frac{da}{a}
g(a) L(a)$ in any positive range of scales and with the coupling
constant $g(a)$ taken from \eqref{glw} is f\/inite.

\section{Conclusion}

In this paper we attempt to construct a theory of scale-dependent
f\/ields $\phi_a(b)$ starting from an action functional that
explicitly depends of these f\/ields, rather than being a local
functional $S[\phi]$ truncated at certain scale $L$ by averaging
over the short-wave f\/luctuations $|k|>\frac{2\pi}{L}$. The
dif\/fe\-rence is that in our theory it is meaningful to consider
the correlation functions $\bra \phi_{a_1}(b_1) \phi_{a_2}(b_2)
\ket$ between dif\/ferent points and dif\/ferent scales. This is
important because any experimental measurement is performed not
exactly in a point, but in a certain vicinity of a point, the size
of which is constrained at least by the uncertainty principle: the
higher momentum is used in the measurement, the smaller is the
vicinity and the stronger is the perturbation caused by the
measurement. Thus we have a strong need do construct a theory in
terms of what is really measured in experiment -- scale-dependent
(wavelet) f\/ields $\phi_a(b)$, rather than in terms of abstract
``no-scale'' functions $\phi(x)$.

Standard RG approach meets this need only partially: it enables to
study the dependence of correlations on separation between points,
$b=b_1-b_2$ in our terms, but not on the typical scale of
interaction. The latter is described in terms of the correlation
length $\xi=\xi(T)$, which has a~universal behavior near the
critical temperature $\xi(T)\sim (T-T_c)^{-\nu}$. The Kadanof\/f
hypothesis that blocks of spins interact exactly as spins
themselves becomes strictly valid only in critical regime
$\xi\to\infty$ at $T\to T_c$. In other cases there are no reasons
to assume that f\/luctuations of dif\/ferent sizes do not interact
to each other and $\bra\phi_{a_1}(b_1) \phi_{a_2}(b_2)\ket=0$ for
$a_1\ne a_2$, instead the inter-scale interaction is expressed by
wavelet diagrams.

As it concerns the criticality, our approach may be close to the
ideas of functional self-similarity (FS)~\cite{KovSh2001}: at the
critical point the blocks of spins can interact with each other in
the same way as primitive spins themselves, but the coupling
constant is scale-covariant, i.e.\ its transformation with
changing of scale depends on the value of coupling constant, but
not on the scale.

For a quantum f\/ield theory models, which arise from  the high
energy physics problems, the Kadanof\/f block-averaging procedure
can not be directly applied. Instead the universal behavior  of
the coupling constant $g=g(a)$ with respect to the scale
transformations $a'=e^\lambda a$ should be considered as a
fundamental symmetry of the physical system. Indeed, the coupling
constant $g(a)$ is a charge of a given particle with respect to a
given interaction measured at a given scale $a$. Since the
experiments can be performed at dif\/ferent scales
$a_0<a_1<a_2<\cdots$, we should expect a universal relation
between the values of charges at dif\/ferent scales, and it should
be an {\it invariant charge}, such that $g_0\equiv
\zeta(a_0,e_0)=\zeta(a_2/a_0,e_2)=\zeta(a_1/a_0,e_1),$ or, since
$e_1=\zeta(a_2/a_1,e_2)$,
\begin{gather}
\zeta(a,g)=\zeta(a/t,\zeta(t,g)). \label{fse}
\end{gather}
The transformation of the charges between dif\/ferent scales
\eqref{fse} has a structure of the Abelian group
$T_{t_1t_2}=T_{t_1}\circ T_{t_2}$. It describes the evolution of
the coupling constant with the change of measurement parameter --
the resolution $a$.

In dif\/ferential form the self-similarity equation \eqref{fse} is
most naturally expressed in the logarithmic coordinate $l = \log
a$:
\[
G(l+\lambda,g) = G(l,G(\lambda,g)),
\]
or
\begin{gather*}
\left( \frac{\d}{\d l} - \beta(g) \frac{\d}{\d g} \right)
G(l,g)=0, \qquad\hbox{where} \quad \beta(g) = \left. \frac{\d
G(\lambda,g)}{\d \lambda} \right|_{\lambda=0}.
%\label{loe}
\end{gather*}
The latter is referred to as the Lie--Ovsyannikov equation.

In general case a quantity $f(l,g)$, which is not an invariant
charge, is not nullif\/ied by the generator of RG transform
$e^{-\lambda \hat R }f(l,g) \ne f(l,g),$ but the generator
\[
\hat R = \left( \frac{\d}{\d l} - \beta(g) \frac{\d}{\d g} \right)
\] determines
the evolution of physical quantities under the scale
transformations.

In our approach we extend the space of functions $\phi(x) \in \lr$
to the space of multiscale (wavelet) f\/ields $\phi_a(b)$ for
which both the  position $b$ and the scale $a$ are considered as
{\em independent} variables. The RG transform therefore becomes
just an Abelian subgroup of the whole symmetry group of the action
$S[\phi_a(b)]$ described above in this paper. The dif\/ference
between our approach and the exact renormalization group approach
(ERG) consists in the fact that ERG generating functional
\begin{gather*}
Z_\Lambda[J] = \int \prod_{|p|\le \Lambda} \cD \phi(p)
e^{-S[\phi;\Lambda]-\int dp J\phi}
%\label{ergz}
\end{gather*}
is just a sum of all f\/luctuations truncated by a cutof\/f
momentum $\Lambda$ in their Fourier representation, and therefore
having certain symmetries related to the rescaling of $\Lambda$ --
the cutof\/f {\em parameter}. In our approach the logarithm of
scale $l=\log a$ becomes an independent variable, so the technique
of Lie symmetry analysis can be applied with respect to $l$ on the
same footing as with respect to the position $x$. In this way the
vector f\/ield that generates arbitrary transformations of the
action $S[\phi_a(b)]$ due to the transformation of dependent and
independent variables takes the form
\begin{gather*}
\hat V = \lambda \left( \frac{\d}{\d l} - \beta(g) \frac{\d}{\d g}
\right) + \xi \frac{\d }{\d x} + \eta \frac{\d }{\d \phi}.
\end{gather*}
We hope that such symmetries could be used to f\/ind solutions of
the wavelet renormalization equation \eqref{phi-loc}.

\subsection*{Acknowledgement}
The author is thankful to the referee for useful comments and
references.

\LastPageEnding

\end{document}